# Analysis of Optical Deshelving in Photon Echo-based Quantum Memories

Byoung S. Ham
*School of Electrical Engineering, and Center for Photon Information Processing
Inha University 253 Yonghyun-dong, Nam-gu, Incheon 402-751, S. Korea*

Storage time extension in photon echoes using optical deshelving via a robust spin state has been investigated for absorption-dependent optical leakage, where an incomplete population transfer, even by a π optical pulse obviates the phase recovery condition of the deshelving. We analyze an optical depth-dependent echo leakage mechanism in the storage time extended photon echoes for the usage of optical deshelving to photon echo-based quantum memories.

Lengthening of photon storage time has been intensively studied recently for quantum memory applications to long-haul quantum communications [1,2]. Optical deshelving has been successfully applied for extended photon storage in modified photon echo schemes, where optical coherence is converted into spin coherence through optical population transfer to an auxiliary spin state [3-6]. Because spin states are much more robust than optical states in an optical medium, lengthened photon storage time can be obtained [7-9]. However, the optical deshelving mechanism by a π optical pulse induces not only population transfer but also phase gain to the transported individual atoms. Thus, for a complete round trip by a consecutive π–π deshelving pulse pair, the total phase shift accumulated becomes π. This π phase shift, however, exactly compensates the rephasing process in photon echoes resulting in no echo generation [10]. To add another π phase shift, an extra rf pulse has been used in between the deshelving pulses [3,11]. Without use of the rf pulse, an all-optical phase recovery condition has been sought [8,10] and experimentally demonstrated [5]. In the phase locked echo, the deshelving pulses require π and 3π pulse areas in series [5,10].

Very recently, the optical deshelving process has been extended for conventional three-pulse photon echoes known as optically locked echoes [12] and atomic frequency comb (AFC) echoes [4]. Unlike phase locked or optically locked echoes, AFC has used identical deshelving pulses, where the phase recovery condition cannot be satisfied. In this AFC deshelving scheme, the accumulated π phase shift should cancel photon echo generation. Surprisingly, the storage time extended photon echoes in AFC have been observed without an additional phase compensation process [4]. Thus, the unexpected photon echo observation in AFC seems contrary to the theory. In this Letter, we investigate the function of deshelving pulses to find out a complete solution to the unexpected photon echo observations in Ref. 4. This understanding sheds light on the usage of photon echoes in ultralong quantum memory applications.

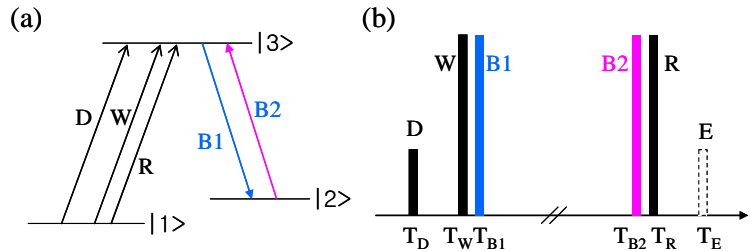

FIG. 1. (a) Partial energy level diagram. (b) Light pulse sequence of (a).

Figure 1(a) shows an energy level diagram of modified photon echoes using deshelving pulses for the purpose of storage time extension. Figure 1(b) shows an optical pulse sequence for Fig. 1(a). The pulses B1 and B2 are for atom deshelving or atom transferring in each state, that is, from the optically excited state |3> to the auxiliary spin state |2> and vice versa. The pulses D, W, and R represent a conventional stimulated photon echo pulse

sequence, where each pulse area Φ is π/2: $\Phi = \int \Omega dt$; Ω is the Rabi frequency of each pulse. In this case inhomogeneously broadened individual atom phase evolutions triggered by D become frozen by W resulting in spectral gratings [13]. When the first deshelving pulse B1 is applied to the state |3>, the excited atoms are transported into an auxiliary spin state |2>, resulting in storage time extension [12].

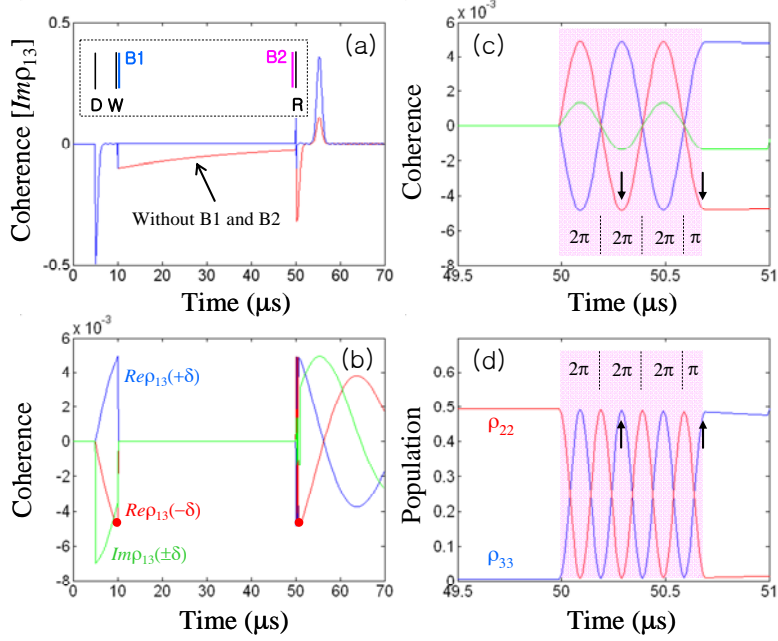

FIG. 2. Optically locked echo using phase locked condition. (a) conventional three-pulse photon echo without B1 and B2 (red) versis optical locked echo with B1 and B2 (blue). Inset: pulse sequence. Pulse duration of D, W, and R: 100 ns corresponding to a π/2 pulse area. B1 and B2: 100 and 300 ns corresponding to π and 3π pulse area, respectively. $T_D$=5 μs; $T_W$=10 μs; $T_{B1}$=10.1 μs; $T_{B2}$=50 μs; $T_R$=51 μs; $\rho_{11}^{(0)}$=1; $\Delta_{inh}$=680 kHz; $\Gamma_{31}=\Gamma_{32}=\gamma_{31}=\gamma_{32}$= 2 kHz; $\Gamma_{21}=\gamma_{21}$=0. (b). Individual atom phase evolution: δ=30 kHz. (c) Expanded figure of (b). Pink shade: B2 pulse. (d) Corresponding figure to (c) for population on each state.

In Fig. 2, time dependent density matrix equations ($i\hbar\dot{\rho} = [H,\rho]$+decay terms) are numerically solved for Fig. 1 to present the phase recovery condition of the deshelving pulses without considering an absorption coefficient or reabsorption of photon echo signals [14,15]. In Fig. 2(a), the enhanced photon echo (blue curve) with π−3π deshelving pulses is compared with a corresponding conventional three-pulse photon echo (red curve), where the enhancement is due to temporary stoppage of the optical population decay process by deshelving atoms into the robust spin state |2>. To understand the phase recovery condition for the deshelving pulses, two symmetrically detuned atoms are taken out from Fig. 2(a) for π−7π deshelving pulse pair (B1 and B2). Here the phase evolution by D is reversed by R via W to generate photon echoes, where the deshelving pulses must be intact on the phase change [see red dots in Fig. 2(b)]. Details of individual atom's phase evolution by the 7π B2 pulse are shown in Fig. 2(c), where the initial phase is retrieved at the pulse area of 3π (first arrow) or 7π (second arrow) of B2: see also Fig. 6 of Ref. 10. This means that the 4π pulse area of B2 induces a 2π phase shift to the atoms due to a double round trip [see Fig. 2(d)]. Figures 2(c) and 2(d) present the π deshelving pulse mechanism, which swaps atom populations between two states with a π/2 phase gain: $\rho_{33} \leftrightarrow \rho_{22}$. Thus, the phase recovery condition of B2 for a fixed B1 pulse (π) is

For B2: (4j−1)π,                   (1-1)
For B1+B2: 4jπ,                    (1-2)

where j is an integer.

For an analytical solution of the deshelving in Fig. 1, we set an absorption parameter η to determine the amount of population transfer. The absorption parameter η is represented by exp(−d), where d is an optical depth (d=α$l$, where α is an absorption coefficient in Beer's law, and $l$ is the length of an optical medium along the light pulse propagation direction). If initial population on the excited state |3> for the B1 deshelving pulse is A, where A is

an entity of optical coherence for photon echoes, then the resulting populations on the excited ($|3\rangle$; $\rho_{33}$) and ground ($|2\rangle$; $\rho_{22}$) states by the $\pi$ B1 deshelving pulse are:

$$\rho_{33} = A(1-\eta)\exp(-t/T_1^{opt}), \quad (2\text{-}1)$$

$$\rho_{22} = A(\eta)\exp(-t/T_1^{spin}). \quad (2\text{-}2)$$

We assume that pulse separation T between B1 and B2 is much shorter than the optical population decay time, $T_1^{opt}$. This assumption holds for most modified photon echo protocols including gradient echoes [6,16,17], AFC echoes [4,18], and phase locked echoes [5,10]. The exponential terms in Eqs. (2) can now be conveniently dropped. Thus, the population $\rho_{33}$ on the excited state $|3\rangle$ by the second deshelving pulse B2 for the first $\pi$ pulse area is simply given by as:

$$\rho_{33} = A[(1-\eta)^2 + \eta^2], \quad (3\text{-}1)$$

where the first term $(1-\eta)^2$ represents no phase shift to the untransferred atoms, while the second term $\eta^2$ is for round trip atoms via state $|2\rangle$ with an accumulated $\pi$ phase shift. This explains why the second deshelving pulse area of B2 should be $3\pi$ for another round trip as discussed in Fig. 2(c), which is three times longer (or stronger) than B1. The auxiliary spin state population $\rho_{22}$ on state $|2\rangle$ by the second deshelving pulse B2 for the corresponding first $\pi$ pulse area is

$$\rho_{22} = A[2\eta(1-\eta)]. \quad (3\text{-}2)$$

For the second $\pi$ pulse area of B2 (total $2\pi$ pulse area), the population $\rho_{33}$ and $\rho_{22}$ are calculated from Eqs. (3):

$$\rho_{33} = A[(1-\eta)^3 + 3\eta^2(1-\eta)], \quad (4\text{-}1)$$

$$\rho_{22} = A[3\eta(1-\eta)^2 + \eta^3], \quad (4\text{-}2)$$

showing no echo contribution due to no phase recovery except the zeroth order, $(1-\eta)^3$. For the third $\pi$ of B2 (total $3\pi$ pulse area), the populations $\rho_{33}$ and $\rho_{22}$ are obtained from Eqs. (4):

$$\rho_{33} = A[(1-\eta)^4 + 6\eta^2(1-\eta)^2 + \eta^4], \quad (5\text{-}1)$$

$$\rho_{22} = A[4\eta(1-\eta)^3 + 4\eta^3(1-\eta)], \quad (5\text{-}2)$$

where the first, second, and third terms of $\rho_{33}$ imply zero, $\pi$, and $2\pi$ phase shift, respectively. Here, the only effective term for the lengthened photon echo generation is the last term, and the phase recovery condition of $2\pi$ is satisfied. Unlike phase locked echo or AFC echo, the first term in Eq. (5-1), however, which adds to the echo generation, cannot be separated due to spatiotemporal overlap.

Owing to the recursive relation between $\rho_{22}$ and $\rho_{33}$ as a function of the B2 pulse area, Table 1 shows the following: $\rho_{ii} = A\sum B_{nm}\eta^n(1-\eta)^{m-n}$, where $B_{nm}=B_{n(m-1)}+B_{(n-1)(m-1)}$, $B_{jj}=1$; $m\geq n$, and ii=33 (22) for even (odd) n's. Here "n" stands for order of $\eta$, while "m" stands for the total pulse area $m\pi$ of the deshelving pulses. The zeroth, 4th, and 8th terms of n imply 0, $2\pi$, and $4\pi$ phase shifts, respectively, and contribute to the photon echoes.

Table 1. Coefficient $B_{mn}$ of $n^{th}$ $\eta$ for deshelving pulse area $m\pi$ for $T \langle\langle T_1^{opt}$.

| n \ m | 0 | 1 | 2 | 3 | 4 | 5 | 6 | 7 | 8 | 9 | 10 |
|---|---|---|---|---|---|---|---|---|---|---|---|
| 0 | 1 | 1 | 1 | 1 | 1 | 1 | 1 | 1 | 1 | 1 | 1 |
| 1 | | 1 | 2 | 3 | 4 | 5 | 6 | 7 | 8 | 9 | 10 |
| 2 | | | 1 | 3 | 6 | 10 | 15 | 21 | 28 | 36 | 45 |
| 3 | | | | 1 | 4 | 10 | 20 | 35 | 56 | 84 | 120 |
| 4 | | | | | 1 | 5 | 15 | 35 | 70 | 126 | 210 |
| 5 | | | | | | 1 | 6 | 21 | 56 | 126 | 252 |
| 6 | | | | | | | 1 | 7 | 28 | 84 | 210 |
| 7 | | | | | | | | 1 | 8 | 36 | 120 |
| 8 | | | | | | | | | 1 | 9 | 45 |
| 9 | | | | | | | | | | 1 | 10 |
| 10 | | | | | | | | | | | 1 |

In Table 1, for m≥4, all pulse areas include the fourth order of $\eta$, which is contrary to the phase recovery condition in Fig. 2. We will hold this matter until we examine the following case.

Before proceeding, we need to briefly mention the AFC echo [4]. Unlike the optically locked echoes [12] in Figs. 1 and 2, the READ pulse R precedes the deshelving pulses in AFC. AFC is an extreme case of the conventional three-pulse photon echoes, where sharp spectral gratings are obtained from repeated weak optical pulse pairs via a complete spontaneous emission process to an auxiliary spin state. To retrieve the stored information, single or multiple pulses (played as the READ pulse R here) are applied. The "trick" of AFC is to use single photons or very weak optical pulses (as R), so that the spectral gratings accumulated with many pulses in advance can be read out several times. The function of the deshelving pulses B1 after R in AFC is to transfer the retrieved (rephased) atoms into state |2>. Unlike optically locked echoes, where the echo position is independent of the deshelving pulse delay, the AFC echo position strongly depends on the delay of B1 from R, similar to the phase locked echo using rephasing process [5,10]. Moreover, the storage time or longest delay time of B2 from B1 is limited by spin inhomogeneous decay, which is much shorter than $T_1^{opt}$ [4,5]. Even though AFC is a kind of three-pulse photon echo, the physics of deshelving follows the phase locked echo as a two-pulse photon echo, which lacks potential for ultralong quantum memories.

To study and visualize the leaked population contribution to the storage time-extended echoes by identical deshelving pulses, we consider a $3\pi-3\pi$ deshelving pulse sequence of B1 and B2 to analyze the echo observation in Ref. 4. For AFC using identical deshelving pulses on a short time scale ($T<<T_1^{opt}$) we start with a $3\pi$ B1 pulse (m=3 in Table 1). Through recursive relation, the final status after a $3\pi$ B2 pulse is (m=6):

$$\rho_{33} = A\left[(1-\eta)^6 + 15\eta^2(1-\eta)^4 + 15\eta^4(1-\eta)^2 + \eta^6\right], \quad (6)$$

where only the third term $15\eta^4(1-\eta)^2$ is effective for photon echoes. Even with a higher order m (wider pulse area of B2), the fourth order of η still exists and increases: see m=10 ($B_{nm}=210$) for B1=B2=$5\pi$ in Table 1. This is due to the leaked population determined by η (discussed in Fig. 3). Thus, an optically shallow medium alleviates the phase recovery condition of the $\pi-3\pi$ deshelving pulse sequence [10].

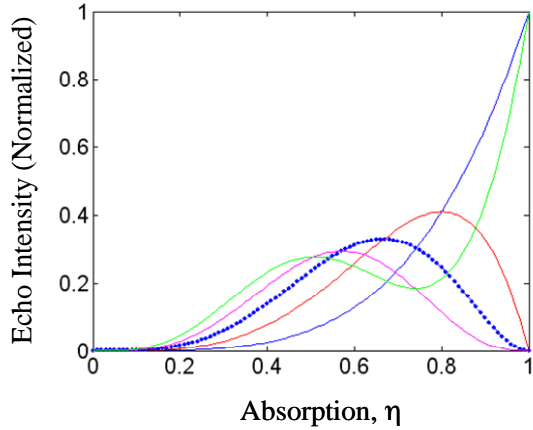

FIG. 3. Excited state population versus absorption. Deshelving pulse areas of B1 and B2: $3\pi-\pi$ (blue); $3\pi-2\pi$ (red); $3\pi-3\pi$ (dotted); $3\pi-4\pi$ (magenta); $3\pi-5\pi$ (green).

Figure 3 shows all extended B2 pulse areas for the fixed $3\pi$ B1 pulse: Eq. (6) is denoted as a dotted curve to analyze Ref. 4. Especially with the 70% absorptive medium (η~0.7; d~1), the photon echo efficiencies are nearly the same for any deshelving pulse area combination, supporting that the optical depth used in Ref. 4 must be dilute; otherwise, no echoes are expected. Although echo efficiency ($\rho_{33}$) is low, the contradictory phenomenon of photon echo observation in Ref. 4 with identical deshelving pulses is explained as an effect of remnant population in a dilute sample. Here, the $3\pi-5\pi$ deshelving pulse sequence shown as green curve satisfying the phase recovery condition (see Eq. 1-2), shows similar pattern to Eq. (5-1) shown as blue curve, if η increases: For the blue curve, the zeroth order in Eq. (5-1) is excluded to satisfy the condition of Ref. 4 as discussed above.

To discuss potential applications of ultralong photon storage, as in the optically locked echoes, we now lengthen the deshelving pulse separation T between B1 and B2: $T_1^{spin} \gg T \gg T_1^{opt}$. In photon echoes, decayed atoms do not contribute photon echoes. Thus, we simply discard $\rho_{33}$ (Eq. 2-1) and use $\rho_{22}$ (Eq. 2-2) for the recursive relation to obtain $\rho_{33}$ satisfying the phase recovery condition ($\pi-3\pi$):

$$\rho_{33} = A\left[3\eta^2(1-\eta)^2 + \eta^4\right], \tag{7}$$

where effective term for photon echoes is the same as Eq. (5-1). For a higher order of B2 (m > 4), however, the $B_{nm}$ of $\eta^4$ decreases by half, explaining why conventional three-pulse photon echo efficiency is always less than 50%. Like short time scale in Table 1, any pulse length of B2 for $T_1^{spin} \gg T \gg T_1^{opt}$ does generate photon echoes due to population leakage in a dilute medium ($\eta$<1). For an optically dense medium ($\eta$~1), however, the phase recovery condition of Eqs. (1) is still satisfied in both cases, because all even-ordered terms of $\eta$ for $\rho_{33}$ contain $(1-\eta)^{m-n}$ except for the highest order given by m. Hence, using a deshelving pulse pair satisfying the phase recovery condition in an optically dense medium ($\eta$~1), and using a phase conjugate scheme [5,7], can give nearly 100% echo efficiency as well as ultralong photon storage time [12].

In conclusion, we investigated the function of deshelving pulses used for photon storage time extension. Due to incomplete population transfer in an optically shallow medium even with a π optical pulse, the phase recovery condition of a deshelving pulse set is alleviated to allow coherence leakage caused by remnant population in the excited state. Although observed photon echoes in Ref. 4 seem to contradict to the theory, the present analysis clearly supports it due to optical leakage in an optically dilute sample. However, the optical leakage-based photon echoes may not be used for quantum memory applications to long haul quantum communications due to less than 50% echo efficiency or shorter storage time. To maximize the echo efficiency, the phase recovery condition of the π−3π deshelving pulse sequence in an optically dense medium should be satisfied.

**Acknowledgments**

This work was supported by the CRI program (Center for Photon Information Processing) of the Korean government (MEST) via National Research Foundation.